\documentclass{10akws-procs9x6}

\def\etal{{\it et al.}}
\def\pt#1{\phantom{#1}}
\newcommand{\bea}{\begin{eqnarray}}
\newcommand{\eea}{\end{eqnarray}}

\begin{document}

\title{TOPICS IN LORENTZ AND CPT VIOLATION}

\author{V.\ ALAN KOSTELECK\'Y}

\address{Physics Department, Indiana University\\
Bloomington, IN 47405, USA}

\begin{abstract}
This talk given at the CPT'10 meeting 
provides a brief introduction to Lorentz and CPT violation
and outlines a few recent developments in the subject. 
\end{abstract}

\bodymatter

\section{Introduction}

The possibility that Lorentz violation might be manifest in nature,
perhaps with attendant CPT violation,
continues to attract attention
from experimentalists and theorists alike. 
In the CPT'07 Proceedings,
I outlined how the triple requirements
of coordinate independence, realism, and generality
lead to the conclusion that
effective field theory 
is the appropriate framework
for studying Lorentz and CPT violation.
The present CPT'10 talk provides some introductory comments 
about this framework.

The comprehensive effective field theory
incorporating General Relativity (GR) and the Standard Model
can be constructed by combining 
all Lorentz-violating operators
together with controlling coefficients
to form observer-invariant terms in the Lagrange density.
This theory is the Standard-Model Extension (SME)
\cite{ck,akgrav}.
A useful limit is the miminal SME,
which restricts operators to mass dimension $d \leq 4$
and is renormalizable in Minkowski spacetime.
Since CPT violation in effective field theory 
comes with Lorentz violation
\cite{owg},
the SME also describes general CPT violation.

Many observable effects arise from the interactions 
of particles with the coefficients,
varying with velocity, spin, flavor, and couplings.
Numerous searches have been performed 
\cite{tables},
but no compelling positive measurement exists to date.
Some intriguing current prospects for signals include,
among others,
oscillations of neutrinos
\cite{neutrinos}
and neutral mesons
\cite{mesons}.

Additional effects occur 
for spontaneous Lorentz violation 
\cite{ksp}
because the coefficients can then fluctuate,
yielding massless Nambu-Goldstone (NG) modes
\cite{ng}
for the broken generators 
\cite{lvng} 
and also massive modes
\cite{lvmm}.
The NG modes can be identified directly with the photon 
in Einstein-Maxwell theory
\cite{lvng},
the graviton in GR
\cite{nggrav}, 
a spin-dependent force
\cite{ahclt}, 
or a spin-independent force
\cite{ngnospin},
or they can generate composite photons
\cite{compphot}
or gravitons
\cite{compgrav}.

\section{Nonminimal terms}

In the full SME with nonminimal terms,
infinitely many possible Lorentz-violating operators 
become candidates for inclusion in the Lagrange density.
As a result,
enumerating these operators
and determining their physical effects becomes challenging.

For operators of arbitrary mass dimension $d$,
a systematic investigation 
has so far been performed only in the photon sector
\cite{km-nonmin}.
This investigation 
studied all operators quadratic in the photon field $A_\mu$,
allowing for arbitrary spacetime derivatives.
The resulting explicit gauge-invariant action reveals 
that the number of Lorentz-violating operators
grows rapidly:
the minimal SME has 4 operators at $d=3$ and 19 at $d=4$,
but 36 nonminimal ones appear at $d=5$, 126 at $d=6$,
and the growth is cubic with $d$ at large $d$.

Each of these numerous operators produces
a distinct Lorentz-violating effect on photon propagation.
In some respects,
the behavior of SME photons is analogous to 
Maxwell photons moving in an anisotropic dispersive crystal.
For example,
Lorentz violation can cause light 
to exhibit mode separation (birefringence),
pulse deformation (dispersion),
and direction dependence (anisotropy).
Certain coefficients for Lorentz violation
can be detected at leading order
by studying propagation in the vacuum,
while others require nonvacuum boundary conditions.
The details of these effects
depend on features of the specific radiation being considered,
such as its frequency, polarization, and direction of travel. 
Surprisingly,
this plethora of new effects is almost unexplored 
in relativity tests.
No dedicated laboratory experiments 
have searched for these behaviors,
and the existing astrophysical tests
are limited to a few comparatively simple cases. 

For coefficients governing leading-order birefringence
in the vacuum,
the most sensitive tests involve polarimetry
of astrophysical sources.
Birefringent effects are controlled 
by the ratio of the wavelength to the source distance,
so the sharpest tests involve polarimetry
of high-frequency radiation propagating 
over cosmological distances.
Although still in its infancy,
the polarimetry of gamma-ray bursts
has already led, 
for example, 
to constraints of order $10^{-32}$ GeV$^{-1}$
on certain operators at $d=5$.

For vacuum-nonbirefringent operators causing dispersion,
interesting tests can be performed
by studying the separation of a propagating pulse.
The sensitivity to the corresponding coefficients
is controlled by the ratio of the pulse separation 
to the source distance. 
For cosmological sources, 
this dispersion-based sensitivity 
is typically many orders of magnitude weaker
than polarimetric measurements,
but nonetheless provides the best access 
to vacuum-nonbirefringent operators. 

Finally, 
for the vast numbers of `vacuum-orthogonal' operators
that produce no leading-order effects 
on photon propagation in the vacuum,
the best option is investigation via laboratory tests.
Typical experiments with resonant cavities and interferometers
produce sensitivities 
given by the ratio of the frequency shift to the frequency. 
Along with studies of astrophysical birefringence and dispersion,
the investigations of these Lorentz-violating effects
on light present an open experimental challenge,
with a real potential for discovery 
in an area that is almost unexplored to date.

\section{Gravity}

The key feature of Special Relativity
is the isotropy of spacetime.
An observable background Lorentz vector or tensor 
implies a spacetime anisotropy of the vacuum 
and hence Lorentz violation.
Similarly,
a key component of GR
is the {\it local} isotropy of spacetime.
Lorentz violation in this context 
can be understood 
as the presence of an observable background vector or tensor
in a local Lorentz frame.

A local Lorentz frame at a given point
is a tangent spacetime to the spacetime manifold.
Since local Lorentz violation
is a property of the tangent spacetime 
rather than the manifold,
the `vierbein formalism' is appropriate 
for studies of local Lorentz violation and gravity.
In this approach,
the vierbein $e_\mu^{\pt{\mu}a}$
implements the conversion from local Lorentz coordinates 
$a$, $b$, $\ldots$
to spacetime manifold coordinates
$\mu$, $\nu$, $\ldots$.

{\it `No-go' result for explicit Lorentz violation.}
The ramifications of these simple observations 
are surprisingly broad.
One powerful result is that explicit Lorentz violation
is incompatible with generic Riemann geometries
and therefore with GR
\cite{akgrav}.
The basic point is that explicit Lorentz violation occurs when 
the background tensors are externally prescribed,
but this is inconsistent with the Bianchi identities
for general Riemann spacetimes.

To illustrate this no-go result,
suppose explicit Lorentz violation appears in the matter sector.
The energy-momentum tensor is then nonconserved
in most spacetimes
and the equations of motion are inconsistent 
with the Bianchi identities,
\bea
0 \equiv D_\mu G^{\mu\nu} = 8 \pi G_N D_\mu T^{\mu\nu} \neq  0
\quad
{(\rm explicit~breaking)}.
\eea
In contrast,
in spontaneous Lorentz violation 
the background tensors are dynamically determined
along with the metric 
and are therefore 
compatible with the spacetime geometry,
\bea
0 \equiv D_\mu G^{\mu\nu} = 8 \pi G_N D_\mu T^{\mu\nu} = 0
\quad
{(\rm spontaneous~breaking)}.
\eea

The no-go result holds also for explicit Lorentz violation
in the gravity sector
and for Riemann-Cartan spacetimes \cite{akgrav}.
In the general case with explicit Lorentz breaking,
imposing consistency with the Bianchi identities 
enforces an additional nondynamical constraint
on the spacetimes solving the theory.
The constraint often forbids any solution,
but in any case it represents at best 
a {\it post hoc} assumption 
slicing the solution spacetimes of the theory.
The no-go result also presents an obstruction 
to reproducing GR from a theory with explicit Lorentz violation,
including theories such as `Lifschitz gravity'
that attempt to generate GR through the running 
of explicit Lorentz-violating couplings.
Gravity theories in which the graviton
arises from spontaneous Lorentz violation
\cite{nggrav}
avoid the no-go result.

{\it Gravitational signals 
from spontaneous Lorentz violation.}
Lorentz violation can occur 
in the pure-gravity and matter-gravity sectors.
The no-go result shows it must be spontaneous,
so the coefficients for Lorentz violation 
must originate as dynamical fields.
Each coefficient field can therefore be written
as the sum of the vacuum coefficient for Lorentz violation
and a fluctuation.
Since the breaking is spontaneous,
the fluctuation includes massless NG modes
and so can affect the dynamics even at low energies.
The problem of solving for these modes and eliminating them 
to recover an effective post-newtonian gravitational theory 
is challenging but has been solved
in both the pure-gravity 
\cite{qbak}
and the matter-gravity 
\cite{ngnospin}
sectors.

Observable effects arise from
Lorentz violation in the gravitational field of the source
and in the trajectory of a test body.
As an example,
the local gravitational acceleration
experienced by a test body near the surface of the Earth
acquires sidereal and annual variations
that can depend on the composition of the test body
and the Earth.
In general,
signals can appear in 
gravimeters 
(free fall and force comparison),
tests of the weak equivalence principle 
(free fall, force comparison, and space based),
exotic matter 
(antihydrogen, higher-generation particles, etc.),
solar-system measurements
(lunar laser ranging, perihelion shift, gyroscopes, etc.),
binary pulsars,
and various photon tests
(Shapiro delay, Doppler shift, 
gravitational redshift, null redshift, etc.).
Also,
a nonzero background torsion can be understood
in terms of certain coefficients for Lorentz violation,
so sensitive constraints on torsion can be obtained
\cite{torsion}. 
The overall prospects for new and improved searches
for gravitational Lorentz violation are excellent.

\section*{Acknowledgments}

This work was supported in part by 
U.S.\ D.o.E.\ grant DE-FG02-91ER40661
and by the Indiana University Center for Spacetime Symmetries.

\end{document}